\documentclass[angeod,manuscript]{copernicus}  %82 pp a 1500
\nolinenumbers
\usepackage{amsmath}
\usepackage{color}

\usepackage{natbib}
\AtEndOfClass{%
  \DeclareRobustCommand*{\vec}[1]
   {\ensuremath{%
     \mathchoice{\mbox{\boldmath$\displaystyle#1$}}
                {\mbox{\boldmath$\textstyle#1$}}
                {\mbox{\boldmath$\scriptstyle#1$}}
                {\mbox{\boldmath$\scriptscriptstyle#1$}}}}}
\renewcommand{\mathbf}[1]{\vec}

\begin{document}

\title{{Entropic Force in MHD, MHD turbulence, and the Entropy-Kinetic Equation 
}}

\author[1]{W. Baumjohann}
\author[2,3]{R. A. Treumann}
\affil[1]{Space Research Institute, Austrian Academy of Sciences, Graz, Austria}
\affil[2]{International Space Science Institute, Bern, Switzerland\footnote{Visiting}}
\affil[3]{Department of Geophysics, Munich University, Germany

\emph{Correspondence to}: Wolfgang.Baumjohann@oeaw.ac.at
}

\runningtitle{Entropic Force}

\runningauthor{Baumjohann and Treumann}

\received{ }
\pubdiscuss{ } %% only important for two-stage journals
\revised{ }
\accepted{ }
\published{ }

%% These dates will be inserted by the Publication Production Office during the typesetting process.

\firstpage{1}

\maketitle

\begin{abstract}
%\noindent\textbf{Abstract}. -- 
Following earlier work, reference is made to the classical entropic force which results from spatially variable disorder, an exclusively repulsive force. In terms of macroscopic variables it is applied to magnetohydrodynamics, causing minor changes on the dispersion of magnetohydrodynamic waves. More important is its effect in magnetohydrodynamic turbulence. Here the entropic force affects the ion inertial-range scales on the electron-dominated perpendicular spectrum causing a steeper than Kolmogorov $\kappa_\perp=-8/3$ spectral slope in agreement with gyro-kinetic simulations. In kinetic theory, inclusion of the ensemble-averaged entropic force weakly modifies the Langmuir wave dispersion. It leads to re-formulation of kinetic theory in terms of Gibbs-Boltzmann-entropy. Liouville's equation in this case becomes an entropy-kinetic equation under Hamiltonian action, with the entropic force-density appearing as an additional force term. This entropy-kinetic equation governs the self-consistent kinetic evolution of the classical entropy including self-modulation.       
\keywords{Entropy Force, MHD, MHD turbulence, Entropy Kinetic Equation}

%\vspace{0.5cm}
%\noindent\textbf{Abstract}.-- 
%\abstract 
%\vspace{1cm}
\end{abstract}
\section{Introduction}

In recent work \citep{treumann2019a} we put forward the notion of a \emph{classical} entropic force as the result of spatial differences in disorder or, if one wants, internal information, the negative of Shannon's external information. We  applied it in passing to the entropy of a Schwarzschild black hole, whose entropy is proportional to its surface and can only increase. This was intuitively natural  to be understood as entropy \citep{bekenstein1972,bekenstein1973}, an insight that had profound consequences in black hole theory. 

In the present note we remain in the realm of less spectacular objects asking for the effects caused by the entropic force in a fluid, more specifically in magnetohydrodynamics (MHD). The entropic force results from the definition of entropy in thermodynamics. Multiplied by temperature $T$ (in energy units) the dimensionless entropy $S$ is encoded in the thermodynamic identity \citep{kittel1980}
\begin{equation}\label{eq-firstlaw}
dU=TdS-PdV+\mu dN +\vec{m}\cdot d\vec{B}
\end{equation}
with $V$ volume, $P$ pressure, $N$ particle number, $\mu$ chemical potential, $\vec{m}$ (elementary) magnetization, $\vec{B}$ magnetic field, as the irrecoverable part of energy which results in internal disorder (heat) or, if one wants, gain of internal information. This suggests that inhomogeneity in disorder is equivalent to the action of a force, which (in contrast to the proposal of a rather different entropic quantum force that has been identified with gravity \citep{verlinde2011}, with consequences in fundamental physics and cosmology) is much more mundane and which we modestly called  \emph{entropy force} \citep{treumann2019a}. In that case the function $TdS(\vec{x})$ is understood as a mechanical potential whose gradient, the spatial differences in disorder (information content) may exert a mechanical force on matter. Here, we adopt the conventional name of an entropic force, however applying it just to classical systems.

As for an example, we first investigate some consequences of this notion on a magnetohydrodynamic fluid, in particular its expected effect on magnetohydrodynamic turbulence. We then turn to the kinetic definition of entropy which, as we demonstrate, also allows the definition of a more fundamental kinetic form of the entropic force. Since this form implies the phase-space distribution we straightly turn to the derivation of a kinetic equation which determines the evolution of entropy in an $N$-particle system. Including the entropic force, we provide its simplified one-particle version which is written in the form of a fundamental classical kinetic equation governing the evolution and self-interaction of entropy in any system subject to an Hamiltonian dynamics. Based on this equation the evolution of (Boltzmann) entropy or its equivalent (Shannon) information it will become possible to develop a general entropic (information) dynamics within the bounds of classical kinetic theory. Here we do not attempt its transformation into the domain of field and quantum theory.    
 
\section{Entropic force density in MHD}
Asking for the effect of entropy it makes little sense to refer to total entropy, which is just an unknown number however large it may become. It accumulates with time with unknown total action in the universe, for instance. Its accumulation may be a driving force, the force that disorder (accumulating heat) exerts. What counts is the local space-time difference $\Delta S$ generated by the considered process. 

For  its explicit determination one needs the entropy as function of the state variables. 

A way to proceed has been opened long ago (by \citet{einstein1905} in his famous work on Brownian motion)  where, from thermodynamic considerations, the Hamiltonian density $h_{E}(\delta\rho)=(\delta\rho/\rho)^2/2\kappa_T$ was obtained. Here $\rho(\vec{x}), T(\vec{x}), \kappa_T(\vec{x})$ are the respective density, temperature, and compressibility and their fluctuations, indicated by the prefix $\delta$,  averaged over a small though sufficiently large volume $\Delta V$. In Einstein's case this Hamiltonian density is given just by the fluctuations of density $\rho$. Remaining in the Gibbs-Boltzmann\footnote{We will be using the name Gibbs-Boltzmann throughout even though we remain in the kinetic frame where the entropy is defined just by the logarithm of the distribution function. Gibbs entropy is the phase-space probability-weighted Boltzmann entropy used in the expectation value of  entropy.} picture of entropy, the probability $P\propto \Omega$ of any macroscopic configuration is proportional to the phase space volume $\Omega=\mathrm{e}^S$. In order to obtain it, one calculates the dimensionless entropy $S_B=\log \Omega$ and its fluctuation induced variation in the non-relativistic approximation, dealing only with classical systems. Then configuration and momentum spaces decouple, with the momentum dependence becoming intrinsic to the entropy, allowing for the use of thermodynamic and statistical mechanical arguments which  later can be extended to kinetic theory.

\subsection{Entropic force potential }
With these remarks in mind  the change $\Delta S$ of the entropy in the chosen spatial volume $\Delta V$ is written
\begin{eqnarray}
\Delta S &=&\!\!\int_{\Delta V}\!\!d\vec{x}\,\rho(\vec{x})\int_0^1\!d\alpha\,\delta s(\vec{x})\cr
&=&\!\!\int _{\Delta V}\!\!d\vec{x}\,\rho(\vec{x})\int_0^1\!d\alpha\,  T^{-1}\big[\delta u+p\,\delta V-\vec{m}\cdot\delta\vec{B}\big]
\end{eqnarray}
Here $\delta s$ is the spatial entropy-density variation which equals the thermodynamic expression in brackets. A magnetization density $\vec{m}$ has been added which contributes to the energy in an external magnetic field $\vec{B}$. In a classical high temperature fluid/plasma the microscopic magnetizations caused by spins and  microscopic atomic orbital motions are of no importance. The magnetic energy density is $B^2/2\mu_0$. However, the macroscopic diamagnetic moment density $\rho\vec{m}=-\rho T/B$ of the particles enters instead, whenever the scale exceeds the gyroscale and the fluctuation induced magnetic moment is adiabatically conserved. In this case the magnetization $\vec{m}$ depends on the magnetic field and the variation also applies to it.  

The clever introduction of the parameter $\alpha$ is due to Einstein who with it took into account all the normalized realizations of the entropy fluctuations which have to be summed over. Evaluating the expression on the right when varying the intensive variables $T(\vec{x})\simeq T(\vec{x},0)+\alpha\delta T,\ p(\vec{x})\simeq p(\vec{x},0)+\alpha\delta p,\ \vec{m}(\vec{x})\simeq\vec{m}(\vec{x},0)+\alpha\delta\vec{m}$, integrating over all realizations $\alpha$, making use of some thermodynamic relations as well as Maxwell-identities (cf., e.g. \citep{landau1994,kittel1980})  yields for the entropy at constant total energy $U$ and volume $V$, up to second order in the fluctuations,
\begin{equation}\label{eq-Sfluct}
\Delta S = -\int_{\Delta V} d\vec{x}\frac{1}{2}\Big[\rho C_v\Big(\frac{\delta T}{T}\Big)^2+\frac{1}{\kappa_TT}\Big(\frac{\delta\rho}{\rho}\Big)^2+\rho\frac{\delta B_\|}{B}-\rho (1-2\cos^2\theta)\Big(\frac{\delta B}{B}\Big)^2\Big]
\end{equation}
where in the scalar product of magnetization density $\vec{m}$ and field $\vec{B}$ we included the projection angle $\theta$, and $\vec{B}\simeq\vec{B}_0$ is the stationary external magnetic field. It should be stressed again that the magnetic terms appear only because in the external magnetic field $\vec{B}_0$ the particles perform a macroscopic gyration with field-dependent diamagnetic magnetic moment $\vec{m}$. We also assume that the number of particles $N$ in the volume is unchanged as usual in the microcanonical case. The integrand of this expression is the variation $\delta s(\vec{x})$ of the spatial entropy density in terms of the variations of density, temperature, and field. $C_v(\vec{x})$ is the specific heat as function of space, and similarly $\kappa_T(\vec{x})=-V^{-1}(\partial V/\partial p)_{T,B}=\rho^{-1}(\partial\rho/\partial p)_{T,B}$ is the  thermal  compressibility. Each of the fluctuations contributes a separate additive term to the entropy. Usually the fluctuations in temperature are less important for the slowly variable temperature. Integration adds just an unimportant constant to the entropy which in most cases can be neglected. Then only the density and magnetic terms remain. We nevertheless provisionally and for completeness retain the temperature fluctuation in the following.

The integrand in the above expression is the entropy density in space. One may note that this density and its spatial variation is immune against local temporally isentropic variations. Multiplication with temperature $T$ then yields the  potential $-\phi_s(\vec{x})$ 
\begin{equation}\label{eq-deltas}
T\delta s(\vec{x})=-\frac{1}{2}\Big[\rho TC_v\Big(\frac{\delta T}{T}\Big)^2+\frac{1}{\kappa_T}\Big(\frac{\delta\rho}{\rho}\Big)^2-\rho T(1-2\cos^2\theta)\Big(\frac{\delta B}{B}\Big)^2+\rho T\frac{\delta B_\|}{B}\Big] = \phi_s(\vec{x})
\end{equation}
whose negative gradient is the mechanical entropic force density 
\begin{equation}
\vec{f}_s(\vec{x})=-\nabla\phi_s(\vec{x})=\frac{1}{2}\nabla\Big[\rho TC_v\Big(\frac{\delta T}{T}\Big)^2+\frac{1}{\kappa_T}\Big(\frac{\delta\rho}{\rho}\Big)^2-\rho T(1-2\cos^2\theta)\Big(\frac{\delta B}{B}\Big)^2+\rho T\frac{\delta B_\|}{B}\Big]
\end{equation}
which is quadratic and thus nonlinear in the temperature and density fluctuations but has a linear term in the fluctuation of the field. This is to be used in any macroscopic fluid description where densities $\rho$ (per mass) are well defined in a given volume $\Delta V$. As argued above, the latter results from the macroscopic magnetization which develops in the presence of a magnetic field $\vec{B}$. It is obvious that, under the assumptions made,  the second order density and temperature fluctuations can be neglected in linear theory. The entropic force enters only in a more elaborate nonlinear treatment. However, in magnetohydrodynamic fluids the last linear term plays a minor role. It depends on the projection of the magnetic field variation onto the external field. In linear theory it comes in only as a compressive low frequency magnetic effect.  

That the entropic force potential is a nonlinear function of the fluctuations was expected, because the kinetic definition of entropy $S=\log f(\vec{x},\vec{v},t)$ through the (dimensionless) distribution function $f$, i.e. the phase-space probability $\Omega\propto f$, is nonlinear. Shannon entropy/information is just the negative of this. It implies that it affects any nonlinear evolution of the fluid where it should be taken into account in the contribution of spatially dependent disorder to the dynamics unless other arguments arise which allow its neglect. Its contribution to nonlinear correlations may become particularly interesting in turbulence theory. 

\subsection{Entropic force-modified mhd-waves}
Let us briefly discuss the possible role of the entropic force on low frequency MHD waves. The only linear term in the entropic force is the magnetic component induced via the orbital magnetic moments of the charges. This should weakly affect low frequency waves the electrically conducting fluid plasma, where at those low frequencies adiabatic conservation of the magnetic moment is natural. These are linear solutions of the magnetohydrodynamic equations. The  fluctuating magnetic field component $\delta B_\|$ along the stationary ambient field $\vec{B}$ is susceptible in Alfv\'en and magnetosonic waves. Adding the entropic force to the linearized momentum equation yields, with $T/m=c_s^2$ the square of sound speed,
\begin{equation}\label{eq-eom}
\frac{\partial\delta\vec{v}}{\partial t}=v_A^2\partial_\|\Big(\frac{\delta\vec{B}_\perp}{B}\Big)-(v_A^2-{\textstyle\frac{1}{2}}c_s^2)\nabla_\perp\Big(\frac{\delta B_\|}{B_0}\Big)-\nabla\Big(\frac{\delta p}{m\rho}\Big)+{\textstyle\frac{1}{2}}c_s^2\partial_\|\Big(\frac{\delta B_\|}{B}\Big)\vec{e}_\|
\end{equation}
(Remember that $\rho$ is just the particle number density.) To this one adds the unchanged induction equation
\begin{equation}
\frac{\partial}{\partial t}\Big(\frac{\delta\vec{B}}{B}\Big)=\partial_\|\delta\vec{v}_{\!\!\perp}-\big(\nabla_\perp\!\!\cdot\delta\vec{v}_\perp\big)\vec{e}_\|
\end{equation}
Combination with continuity and equation of state leads to the slightly modified dispersion relation of magnetosonic waves
\begin{equation}
\Big[\omega^2-k_\|^2v_A^2-k_\perp^2\big(v_A^2+{\textstyle\frac{1}{2}}c_s^2\big)\Big]\big(\omega^2-c_s^2k_\|^2\big)-{\textstyle\frac{3}{2}}c_s^4k_\|^2k_\perp^2=0
\end{equation}
The isentropic alfv\'enic dispersion relation $\omega^2-k_\|^2v_A^2=0$ remains of course unchanged. The entropic force just causes a minor change in the  speed $c_{ms}$ of infinitesimal-amplitude magnetosonic waves in magnetohydrodynamics and introduces a factor $\frac{3}{2}$ in the last term. These lesser modifications are caused by the linear term in the entropic force when accounting for the induced magnetic moments $m=-T/B$ on the particles in the presence of a sufficiently strong magnetic field $\vec{B}$. It is clear that in sufficiently strong magnetic fields with $v_A^2\gg c_s^2$, the usual case in the solar wind or near Earth space physics, the entropic contribution is widely negligible as neither sound nor  sound velocity play any susceptible role. However, in  weakly magnetized fluid dynamics their contribution is not negligible. 

Otherwise the entropic force is a nonlinear force. Thus, in the nonlinear evolution and theory of magnetohydrodynamic waves it is expected to play some more important role this time then not through the magnetic fluctuations but the density modulation. Thus, when going nonlinear, its neglect of the nonlinear terms in the entropy potential can hardly be justified how difficult its inclusion may ever become. 

In general, however, when considering high frequency waves in plasmas the entropic force will have to be included either as an additional fluid force through the above potential or, on the kinetic level, expressing the density and temperature fluctuations through their moments with respect to the kinetic distribution function. In that case the magnetic contribution at high frequencies vanishes as the magnetic moments of the charges will not be conserved anymore. Any magnetic wave component then contributes as usual solely through electrodynamic coupling. 

\subsection{Entropic force in MHD Turbulence}
The most interesting contribution of the entropic force to magnetohydrodynamics is expected to occur in turbulence theory. Turbulence is a prominent nonlinear phenomenon \citep{biskamp2003}. One expects that the entropic force will intervene ever when turbulence evolves and turbulent disorder arises, heat is exchanged, and the entropy exhibits local fluctuations. 

Since in turbulence theory the correlation moments dominate, the intrinsically nonlinear entropy potential necessarily contributes. In a fluid system, the fluctuations are given by (\ref{eq-Sfluct}). Turbulence evolves on different scales. Measurements provide statistical averages of the turbulent energy transport through the scales. Theory thus deals with the energy transport equation which describes the flow of energy density and its distribution onto the sequence of decreasing scales. Accounting in addition for the entropic force implies completing the energy-density flow with entropy-density flow  $\vec{v} Ts(\vec{x})=-\vec{v}\phi_s(\vec{x})$ in the energy transport equation, causing  additional complications which so far have been neglected. 

Turbulence theory in magnetohydrodynamics is based on the full system of dissipative low-frequency hydrodynamic equations including the Lorentz-force and the electrodynamic fields.  Including the entropic force the equation of motion of  turbulent magnetohydrodynamic fluctuations becomes 
\begin{eqnarray}\label{eq-eom}
\frac{d\delta\vec{v}}{dt}&=& \frac{1}{\rho}\Big\{-\nabla\Big[P_\mathit{th}+\frac{\delta B^2}{2\mu_0}\Big(1-\frac{v_\mathit{th}^2}{2v_A^2}(1-2\cos^2\theta)\Big)+\frac{\vec{B}\cdot\delta\vec{B}}{\mu_0}\Big(1-\frac{v_\mathit{th}^2}{4v_A^2}\Big)+\rho TC_v\Big(\frac{\delta T}{T}\Big)^2\cr
&&-\frac{1}{2\kappa_T}\Big(\frac{\delta\rho}{\rho}\Big)^2\Big]
+\frac{1}{\mu_0}(\vec{B}+\delta\vec{B})\cdot\nabla\delta\vec{B}\Big\}+\nu\nabla^2\delta\vec{v}+{\textstyle\frac{2}{3}}\nu\nabla\nabla\cdot\delta\vec{v}
\end{eqnarray}
 which is to be supplemented by the low frequency electrodynamic equations. We retain the temperature variations for completeness, because they contribute to higher order entropy-modified transport. Here they contribute just a constant which drops out when applying the gradient. Similarly, the heat or energy transport equation would have to be rewritten to include the above defined entropy-density flow. This becomes necessary when expressing the correlations and developing the full renormalization group theory of turbulence in configuration space \citep{forster1977} as, for instance, has been initiated in \citep{verma2004}. Since  we will not refer to it below, we do not write it down at this place. 

Assume for simplicity an ideal gas equation of state $P_\mathit{th}=\rho T$, with $T=\frac{1}{2}v_\mathit{th}^2$ (remember that all quantities are understood here as per mass). The coefficient $\nu$  is the kinematic viscosity.  From the last expression it is seen that the magnetic terms simply add to magnetic pressure. 

An interesting problem arises with the density fluctuations $\delta\rho$. Before discussing it we rewrite the total pressure term in the last equation as
\begin{equation}
P_\mathit{tot}\equiv P_\mathit{th}+\frac{\vec{B}\cdot\delta\vec{B}}{\mu_0}\Big(1-\frac{v_\mathit{th}^2}{4v_A^2}\Big)+\frac{(\delta B)^2}{2\mu_0}\Big[1-\frac{v_\mathit{th}^2}{2v_A^2}(1-2\cos^2\theta)\Big]+\rho TC_v\Big(\frac{\delta T}{T}\Big)^2-\frac{1}{2\kappa_T}\Big(\frac{\delta\rho}{\rho}\Big)^2
\end{equation}
The entropic force causes a modification of the magnetic pressure. It apparently also adds a contribution from the turbulent density. However, this density contribution contains the compressibility $\kappa_T=-V^{-1}(\partial V/\partial P_\mathit{th})_T$, which is a transport coefficient, indicating that the corresponding term is dissipative and should be related to the fluctuation $\delta\vec{v}$ of the velocity instead  the pressure. 

To see this more clearly, consider the Lorentz force electric field $\vec{E}=-\vec{v}\times\vec{B}-\eta\vec{j}$ where $\vec{j}=
\nabla\times\vec{B}/\mu_0$ is the low-frequency electric current, and $\eta$ resistivity, assumed constant here.\footnote{Constancy of $\eta$ in turbulence is not given a priori. Nonlinear reaction of turbulence on transport is a normal process which, however, is of higher order in the fluctuations and needs not to be considered here. It should be accounted for in a renormalization group theory of turbulence.} We first note that stationary turbulence implies that $\nabla\times\delta\vec{E}=0$ which, from Faraday's law and incompressible turbulence  $\nabla\cdot\delta\vec{v}=0$ and vanishing average flow speed $\langle\vec{v}\rangle=0$, for instance, yields that to lowest order  
\begin{equation}
\vec{B}\cdot\nabla\delta\vec{v}=\frac{\eta}{\mu_0}\nabla^2\Big(\frac{\delta\vec{B}}{B}\Big)
\end{equation}
which relates the magnetic fluctuations in incompressible turbulence to the velocity fluctuation field with magnetic diffusion coefficient $\eta/\mu_0$, an expression  used to obtain the turbulent magnetic density when the stationary turbulent velocity field density is known. As usually, however, in collisionless ideal flows with $\eta=0$ this vanishes identically.

Density fluctuations can, in magnetohydrodynamic turbulence theory, mostly be neglected at low frequencies for the assumption of quasi-neutrality. Continuity of flow $\delta\rho/\rho\sim\delta v/\langle v\rangle$ suggests that the  spectrum of density fluctuations along the mean flow $\langle\vec{v}\rangle\neq0$ should resemble that of the velocity fluctuations. In the absence of a mean flow this relation becomes obsolete, though distinctions have been made between large and small scale turbulence, when the large scales represent mean flows for the small scales \citep{tennekes1975}. Nevertheless, in spite of the quasineutrality which usually is considered valid to hold, with the exception of the high frequency range on the respective electron plasma $\omega_e^{-1}$ or cyclotron $\omega_{ce}^{-1}$ time scales, various observations of turbulence in the solar wind (cf., e.g. \citep{podesta2007,podesta2009})  have identified turbulent density spectra in the solar wind. An explanation of the modulated spectral slope has been given \citep{treumann2019b} based on divergence of the Lorentz force. There is good reason to assume that this is the case in the ion inertial range where the ions drop out from being magnetized just following their inertia, and the electron behaviour is quite different. Currents are then carried solely by the remaining magnetized electrons. These scales are subject to kinetic Alfv\'en wave generation, propagation and dissipation, and it is, presumably even much more important, subject to reconnection in electron-scale current filaments, the most probable mechanism of kind of violent (spontaneous, in contrast to moderate weakly-turbulent wave-induced) dissipation of mechanical energy in turbulence \citep{treumann2013,treumann2015} far above the molecular scale. In a completely ionized collisionless fluid on scales $L\gg\lambda_D$ exceeding the Debye length, all electrons participate in the divergence of the induced electric Lorentz force field $\vec{E}$, which is equivalent to an electronic charge density on those scales. 

Taking the divergence of the Lorentz field $\vec{E}=-\vec{v}\times\vec{B}-(\eta/\mu_0)\vec{j}$ eliminates the current $\nabla\cdot\vec{j}=0$. Poisson's equation gives the relation between charge density, field, and velocity fluctuations 
\begin{eqnarray}\label{eq-density-mhd}
\frac{eq\delta\rho}{\epsilon_0}&=&-\big(\vec{B}+\delta\vec{B}\big)\cdot\big(\nabla\times\delta\vec{v}\big)+\big(\langle\vec{v}\rangle+\delta\vec{v}\big)\cdot\nabla\times\delta\vec{B}\cr &\longrightarrow&\ -B[\vec{z}\cdot(\nabla\times\delta\vec{v})] =-B(\nabla\times\delta\vec{v})_\|
\end{eqnarray}
with  electron charge $-e$, average velocity $\langle\vec{v}\rangle$, and $\vec{z}$ the unit vector along the ambient magnetic field $\vec{B}$. Still, we cautiously included a fraction $q\lesssim1$ of electrons here in order to maintain the freedom that a fraction $1-q$ of high energy electrons with large gyroradii exceeding the ion inertial length escape the Lorentz force on these scales. Only the linear term needs to be retained, here, thus neglecting the average flow velocity $\langle\vec{v}\rangle$ and the mixed terms in the density expression (\ref{eq-density-mhd}). 
Applying the gradient to the entropic force potential $\phi_s(\vec{x})$ one obtains for the contribution of the density fluctuation
\begin{equation}
\frac{1}{2\kappa_T\rho}\nabla\Big(\frac{\delta\rho}{\rho}\Big)^{\!2}= \frac{1}{q^2\omega_i^2}\Big(\frac{v_A}{c}\Big)^{\!2}\Big(\frac{\partial P_\mathit{th}}{\partial\rho}\Big)_\mathit{\!\!T\!,\!B}\big(\nabla\times\delta\vec{v}\big)_\|\big[\nabla\big(\nabla\times\delta\vec{v}\big)_\|\big]
\end{equation}
where we inserted for the compressibility $\kappa_T=\rho^{-1}(\partial\rho/\partial P_\mathit{th})_{T,B}$.  This expression is  of the form of the last viscous term in Eq. (\ref{eq-eom}) with velocity dependent factor which has the correct dimension of a viscosity. Only the field-parallel component of the vector product appears here both in the first and second order differentials. Thus the whole expression maintains the scalar properties while applying to the perpendicular turbulent velocity fluctuations. Note that this term does not vanish even in incompressible turbulence $\nabla\cdot\delta\vec{v}=0$ as it affects only the perpendicular velocity component $\delta\vec{v}_\perp$. It will always be present whenever velocity fluctuations evolve coupling to density fluctuations $\delta\rho$, here through Faraday's law respectively the divergence of the Lorentz force. Comparing with the above equation of motion in viscous magnetohydrodynamics, its factor formally defines a second turbulent entropic viscosity through
\begin{equation}
\nu_\mathit{S}(\delta\vec{v})=\frac{3}{2q^2\omega_i^2}\Big(\frac{v_A}{c}\Big)^{\!2}\Big(\frac{\partial P_\mathit{th}}{\partial\rho}\Big)_\mathit{\!\!T\!,\!B}\big|\nabla\times\delta\vec{v}\big|_\| \ \propto\ \nu_{S0}\,\big|\nabla\times\delta\vec{v}\big|_\|
\end{equation}
This viscosity is not constant but depends on the (transverse) velocity fluctuations $\delta\vec{v}_\perp$. Up to a dimensional constant it however defines a velocity independent viscosity $\nu_{S0}$. It thus through the velocity depends on the scale. In collisionless magnetohydrodynamics $\nu=\eta=0$ this will actually become the only and thus non-negligible viscous term, produced solely by the entropic force, i.e. by space dependent turbulent disorder respectively internal heat manifesting itself in density fluctuations. Still it needs to be represented through the averaged fluctuations in density and/or velocity. Fourier expanding according to
\begin{equation}
\delta\vec{v}(\vec{k})=(2\pi L)^{-d}\int d^d{x}\,\delta\vec{v}(\vec{x})\,\mathrm{e}^{-i\vec{k}\cdot\vec{x}}
\end{equation}
 with $d$ dimension and $L$ length, we have, with $\delta(k_\|)$ accounting for the absence of the parallel wavenumber,
\begin{equation}\label{eq-avdeltarho}
\langle\big|\delta\rho^2(\vec{x})\big|\rangle \sim \langle\big|(\nabla\times\delta\vec{v})_\|\big|^2\rangle =\int d\vec{k}\,k_\perp^2\mathcal{S}_\perp(k)\delta(k_\|)
\end{equation}
Here $\mathcal{S}_\perp(k)$ is the spectral energy density of the perpendicular velocity fluctuations from 
$\int d\vec{k}\,S(k)=\frac{1}{2}\langle|\delta\vec{v}|^2\rangle$. With $\nu_{S0}$ the above dimensional factor, we obtain for the expectation value of the viscosity
\begin{equation}
\langle|\nu_S|^2\rangle=\nu_{S0}^2\int d\vec{k}\,k_\perp^2\mathcal{S}_\perp(k)\delta(k_\|)
\end{equation}
This is proportional to the spectral density in the transverse velocity fluctuations and clearly scale-dependent. It folds into the diffusive term in the equation of motion. We may reasonably assume that the spectral interaction is local, depending only on the differences $\vec{k}'-\vec{k}$ in wave number. Then the right hand side of this equation becomes proportional to  the expectation value of the dissipated energy $\varepsilon_{S}$ per unit of time, which identifies it as
\begin{equation}\label{eq-endissip}
\varepsilon_{S}\ = \nu_{S0} \int d\vec{k}\, k^2_\perp\mathcal{S}(k_\perp)\delta(k_\|)\ = \frac{\langle|\nu_S|^2\rangle}{\nu_{S0}}
\end{equation}
We are less interested neither in the turbulent viscosity itself, which by the above expression for $\nu_{S0}$ is not a large number, nor in the amount of energy dissipated due to the presence of the entropic force and its dissipative potential. Our interest is in any possible effect on the shape of the wave number spectrum of turbulence $\mathcal{S}(\vec{k})$. To infer about this, a simple dimensional analysis suffices when following the adopted procedure, assuming that the energy $\varepsilon_S$ is dissipated at a certain dissipation scale $\ell_d$ and that possibly an inertial range exists, where the viscosity $\nu_{S}$ does not depend on scale and dissipation consists merely in the nonlinear transport of energy from large to small scales or vice versa, according to the action of the entropy, the second law, and the entropy-generated viscosity.  If such an inertial region exists at all, then the viscosity $\nu_{S}$  in this region should not have any influence on the shape of the spectrum, which means that the spectrum should be self-similar exhibiting some constant spectral slope.  

Adopting this approach, Eq. (\ref{eq-endissip}) gives dimensionally when multiplying with powers of the dissipation length $\ell_d$ and identifying the dissipated energy $\varepsilon_S=\varepsilon_{Sd}$ per unit time with the energy dissipated at the dissipation scale
\begin{equation}
\mathcal{S}(k_\perp)\sim \frac{\varepsilon_{Sd}}{\nu_{S0}}\ell_d^{2+d}(k_\perp\ell_d)^\kappa F(k_\perp\ell_d)
\end{equation}
Here $F(k_\perp\ell_d)$ is the scale-free function that cares for self-similarity. Dimensionally we have $[\varepsilon_S]\sim\mathrm{L}^2/\mathrm{T}^3$ and $[\nu_{S0}]\sim\mathrm{L}^2/\mathrm{T}$, and after eliminating the time dimensions in the energy flow and viscosity, also for $\mathrm{L}\sim\ell_d=(\nu_{S0}^3/\varepsilon_{Sd})^{1/4}$. Inserted into the last equation and requiring that the power of the viscosity vanishes (because in the inertial-scale region viscous dissipation has no effect on the spectral shape), we obtain for the inertial slope of the spectrum 
\begin{equation}
\kappa=-{\textstyle\frac{1}{3}}(2+3d)
\end{equation}

In one dimension with $d=1$ this would recover Kolmogorov's spectrum. However, the entropic viscosity in our approach works exclusively in the two perpendicular dimensions only as prescribed by the Lorentz force. Thus we have $d=2$ obtaining a spectral shape of the perpendicular turbulent velocity spectrum
\begin{equation}
\mathcal{S}(k_\perp)\sim k_\perp^{-8/3}F(k_\perp\ell_d)
\end{equation}

This perpendicular spectrum is steeper than Kolmogorov being close to $\mathcal{S}(k_\perp)\sim k_\perp^{-3}$. It should, however, be noted that it applies to the ion inertial range scale region $\lambda_e<k_\perp^{-1}<\lambda_i$ (with $\lambda_{e,i}=c/\omega_{e,i}$ the skin depths of electrons and ions), where electrons and ions magnetically decouple. Here, the electrons still provide a dissipationless non-viscous scale-invariant spectral energy flux from large to small scales while the ions do not anymore participate. They already act dissipating their turbulent energy contributed through the turbulent viscosity $\nu_{S0}$. The scale-free spectrum thus necessarily should become anisotropic here and decay at a steeper rate in wave numbers, an effect due to the entropic force in otherwise completely collisionless magnetohydrodynamics.  

Perpendicular spectra of slope $\kappa_\perp\approx 2.7$ very close to the above prediction have indeed been obtained in numerical gyro-kinetic simulations of solar wind turbulence in the ion inertial scale range \citep{howes2011} where they have tentatively been attributed to the sole dissipative action of kinetic Alfv\'en waves. Here they arise from fundamental theory without reference to any waves or instabilities as the direct signature of the entropic force respectively the action of entropy in the ion inertial scale range. It should be noted in addition that the entire theory presented here is done under the assumption of absence of any viscosity other than that introduced by the entropic force. 

One may again repeat  that independence of slope from viscosity does not imply $\nu_{S0}=0$. It just means that the spectral slope is not deformed by the constant viscosity. It is instead maintained by its uniform action transporting energy from larger to shorter scales which is nothing else but just sort of some dissipation which acts in resolution of scales into smaller elements though not transforming energy into heat. However, outside the entropic inertial range the entropic viscosity acts like a normal viscosity, transforms energy into heat and thus will necessarily deform the spectrum. Its mere presence, if at all remarkable, outside the entropic inertial range, for instance in the Kolmogorov or Kraichnan ranges, should cause deviations from the inertial spectral slope and possibly lead to a reduction of the extension of the inertial range, an effect observed both in nature and simulations. 

Does one expect a similar though separate effect leading to a magnetic spectrally different slope when considering the contribution of magnetic fluctuations to the entropic force? The answer is no, whenever the turbulence is stationary with vanishing average flow $\langle\vec{v}\rangle=0$. Then under stationary conditions and incompressibility Faraday's law does not couple the additional magnetic entropy contribution to the turbulent velocity fluctuations. Thus the magnetic spectrum should simply mirror the velocity spectrum. This might supposedly change in flowing turbulence (like the solar wind) where Faraday's law imposes a relation between the divergence of the perpendicular velocity  and the  parallel field fluctuations
\begin{equation}
(\langle\vec{v}\rangle\cdot\nabla)\delta B_\|=B\nabla_\perp\cdot\delta\vec{v}_\perp
\end{equation}
which spectrally translates into
\begin{equation}
k_\perp^2\mathcal{S}(k)=\langle v^2\rangle k_v^2\mathcal{S}_{B_\|}(k)/B^2
\end{equation}
where $k_v$ is the wave vector of the magnetic fluctuations along the average flow. As usually, this suggests that the spectrum of parallel magnetic fluctuations simply translates into that of (perpendicular) velocity fluctuations, just depending on the angle of flow and constant ratio $\langle v^2\rangle/v_A^2$. The average over the angle with respect to the direction of $k_v$ merely gives a number while maintaining the spectral slope. 

The result of this section which says that in collisionless MHD with $\nu=0$ and absence of any collisional dissipation on scales far above the molecular, is that the entropic force itself provides a finite viscosity $\nu_{S0}$. However small, it causes a feedback of the turbulent fluctuations on turbulence whose nature through it becomes viscous and dissipative.  Generation of this viscosity should be a general phenomenon in MHD though its effect might mostly be not remarkable or within the range of uncertainty. In the presence of any otherwise generated anomalous sufficiently large viscosity, its effect on the shape of the inertial range spectrum will probably become buried below the general spectral trend. The inertial range will remain determined by kind of Kolmogorov spectral decay as is mostly observed in the solar wind where it is determined up to the ion inertial scale by  fluid turbulence. When entering the ion inertial scale, however, the turbulence should  become anisotropic because the inertial range spectrum is continued only into the perpendicular spectrum where, in the absence of any otherwise generated dissipation, the spectrum decays more steeply than Kolmogorov until the turbulence reaches the electron inertial scale where it should abruptly end in dissipation.   

\section{General entropic force in kinetic theory}
After this brief excursion into magnetohydrodynamics and turbulence, we return to the main problem of the definition of the entropic force. On the kinetic level, the entropy has been  defined by Gibbs-Boltzmann in its famous logarithmic form $S_B$ as given above. Assume there is a classical probability distribution in phase space $f(t,\vec{v},\vec{x})$ which obeys the Liouville equation 
\begin{equation}\label{eq-liou-1}
\partial_t f(t,\vec{x},\vec{v})+[\mathcal{H},f(t,\vec{x},\vec{v})]_P = \mathcal{C}
\end{equation}
where $\mathcal{H}$ is the Hamiltonian of the system from that the equations of motion in momentum/velocity and configuration space can be derived, and $\mathcal{C}$ some source (or collision) term on the right which is positive if acting as source, negative if dissipating and causing losses, and the brackets are Poisson brackets (as indicated by the subscript $P$) defining the particle dynamics. We are dealing here for simplicity with particles not with fields. The Liouville equation has to be completed with the equations which describe the interaction between the particles in the volume, and these interactions are mediated by fields if not simply by collisions. Then, in terms of the phase-space probability distribution, the kinetic Gibbs-Boltzmann entropy is defined as $S_B=\log f$. The total entropy of the entire system which would become involved into the thermodynamic potentials is of course the integral all over phase space 
\begin{equation}
\langle\, S\,\rangle=\int d\vec{x}d\vec{v}\,f(\vec{x},\vec{v})\log f(\vec{x},\vec{v})
\end{equation}
as the average moment of the entropy (or in other terms its statistical expectation value). Here $\log f=S_B$ is again Gibbs-Boltzmann's  microcanonical entropy, which is to be multiplied by $f(\vec{x},\vec{v})$ and integrated in order to provide the total entropy. The momentum space integral defines the entropy density in real space
\begin{equation}
s(\vec{x})=\int d\vec{v}\, f(\vec{x},\vec{v})\log f(\vec{x},\vec{v})
\end{equation}
Assume that the temperature $T$ of the system is known. Multiplying by $T$ and applying the spatial gradient we obtain the microcanonical Gibbs-Boltzmann entropic force density
\begin{equation}
\vec{f}(\vec{x})=-\nabla [Ts(\vec{x})]
\end{equation}
In this expression the temperature can still be a function of space and time. From this definition follows for the entropic force-potential  
\begin{equation}
\phi_s(\vec{x})=Ts(\vec{x})
\end{equation}
which is a mechanical potential that is to be added to the potential either in the Hamiltonian $\mathcal{H}$ or the equivalent Lagrangean $\mathcal{L}$. Its sign depends upon the sign of the logarithm. Since $f < 1$ usually is a normalized probability density, this sign is negative, and the entropic force is repulsive. It acts against gravity causing spatial expansion of any inhomogeneities in disorder respectively expansion of internal information.  

The important difference between this and any other force potential is that the entropy potential is defined solely through the phase space density itself, not by other external fields. More precisely, it is defined through its integral over the chosen volume $\Delta V_{\vec{v}}$ in momentum space. In this sense it describes the self-interaction and feedback of the phase-space density on itself through the changes of the phase-space density induced by the fields which determine the particle dynamics. 

Clearly this self-interaction is highly non-linear, and it becomes even more complicated by the fact that the temperature itself is determined as $T=(\partial_S U)_V$, the derivative of the total energy with respect to the entropy. However, even in the case of global equilibrium with constant overall temperature $T=$ const, which does still allow for local non-equilibria, the intrinsic nonlinearities remain. Through them, the entropy density of the system affects itself, something well known from field theory for instance. It moreover says that the \emph{entropy}, at least under certain so far unspecified conditions, \emph{behaves as its proper source}. 

The above definition of the entropic force density through the distribution function and integral over the momentum space volume identifies the entropic force density and its potential as independent of the momentum and therefore as a conservative force. However, its definition through an integral over momentum space turns the Liouville equation (\ref{eq-liou-1}) into an integro-differential equation for the phase space distribution
\begin{equation}\label{eq-liou-2}
\partial_t f(t,\vec{x},\vec{v})+[\mathcal{H}_0,f(t,\vec{x},\vec{v})]_P -\nabla\Big(T\int d\vec{v}\,f(\vec{x},\vec{v})\log f(\vec{x},\vec{v})\Big)\cdot\partial_{\vec{v}}f(\vec{x},\vec{v},t)= \mathcal{C}
\end{equation}
where $\mathcal{H}_0=\mathcal{H}-\phi_s$ is the (entropic potential)-reduced Hamiltonian. One may note that the term in round brackets is independent on momentum/velocity as this dependence is integrated over, such that the momentum gradient can be put in front. For constant temperature this simplifies to become
\begin{equation}\label{eq-liou-3}
\partial_t f(t,\vec{x},\vec{v})+[\mathcal{H}_0,f(t,\vec{x},\vec{v})]_P -\partial_{\vec{v}}f(\vec{x},\vec{v},t)\cdot \int d\vec{v}\ [T\,\nabla f(\vec{x},\vec{v},t)]\big(1+\log f(\vec{x},\vec{v},t)\big) = \mathcal{C}
\end{equation}
These equations show the self-interaction of the phase space distribution through its self-generated entropy.  In all cases when the distribution function develops an inhomogeneity there arises an unavoidable collisionless reaction of the phase space distribution on its own evolution, with nonlinear self-interaction term. It results from the nonlinearity of the entropy and thus enters quite naturally. However, due to the nature of the entropy, the local self-reaction of the distribution function through the entropy is a non-local effect in momentum space as it involves a finite momentum space volume which has to be integrated over, whereas in configuration space it is local in the infinitesimal environment of the local volume as selected by the spatial gradient.

We note that any of these equations can be taken as an evolution equation of entropy where the entropy turns out as its own source. This evolution equation of the entropy can be obtained when multiplying the Liouville equation with the logarithm of the distribution function.    

In spite of the fact that the entropy regulates itself, the last expression (\ref{eq-liou-3}) is interesting when applied in linear theory. Only the last term needs to be considered when introducing $f=f_0+\delta f$. Neglecting all nonlinear terms and observing that under the integral such products like $f_0\log f_0$ when integrated just produce the constant global entropy which vanishes when $\nabla$ is applied to it. The only two terms remaining, one the velocity integral over the fluctuation of the distribution function which thus just produces the fluctuating number density $\int d\vec{v}\,\delta\!f(\vec{x},\vec{p},t)=\delta \rho(\vec{x})$. The other remaining term is mixed, with integrand $(f_0\log f_0)\delta\!f/f_0$ where the term in brackets is the overall density of the entropy. Partial integration then lets the stationary term vanish. In the remaining integral then the density of entropy $s_0$ is independent of velocity and can be taken out of the integration leaving a total derivative of $\delta f$  unter the integral. Hence this term vanishes as well and thus does not contribute linearly. In linear approximation on the kinetic level, the entropic force then adds only the additional force term
\begin{equation}
-\nabla( T\delta \rho/\rho_0)\cdot\partial_{\vec{v}}f_0(\vec{v})
\end{equation}
to the linear kinetic equation. The linearized kinetic equation becomes finally 
\begin{eqnarray}
\partial_t\delta\!f + [\mathcal{H}_0,\delta\!f]_P&=&\nabla\big(T\delta\rho[\delta\!f]/\rho_0\big)\cdot\partial_{\vec{v}}f_0(\vec{v})\\
\delta\rho(\vec{x},t)&=&\int_{\Delta V_{\vec{v}}}d\vec{v}\, \delta f(\vec{x},\vec{v},t)\label{eq-dens-fluct}
\end{eqnarray}

The spatial gradient of the density fluctuation is multiplied by the momentum space gradient of the equilibrium distribution function $f_0(\vec{v})$. Here the density fluctuation $\delta\rho$ arises from the momentum space integral of the fluctuation of the distribution function $\delta\!f$. In any particular problem the density fluctuation is to be expressed through the field equations which, in the electrodynamic case are the Maxwell equations where the density appears in Poisson's law and becomes re-expressed through its above integral form which thus can be used directly. 

As for an example, let us briefly consider  electrostatic electron (Langmuir) waves in one dimension. Inclusion of the entropic force modifies the linearized Fourier-transformed Vlasov equation to become 
\begin{equation}
-i(\omega-kv)\delta f_{k\omega} = \big[(e/m)\delta E_{k\omega}+ikT(\delta\rho_{k\omega}/\rho_0)\big]\partial_vf_0(v)
\end{equation}
where we reintroduced the electron mass $m$. This is to be completed by Poisson's law for the Fourier transformed electric field $E_{k\omega}$ as
\begin{equation}
ik\delta E_{k\omega}=-(e/\epsilon_0)\delta\rho_{k\omega} 
\end{equation}
which is used to eliminate the density fluctuation $\delta\rho_{k\omega}$ to yield an expression for $\delta f_{k\omega}$. Inserting this into (\ref{eq-dens-fluct}) to express $\delta\rho_{k\omega}$ and then again into Poisson's law eliminates the electric field from both sides and yields the slightly modified one-dimensional complex dispersion relation of electrostatic unmagnetized-electron plasma (Langmuir) waves with plasma frequency $\omega_e$: 
\begin{equation}
\mathcal{D}(k,\omega)=1-\frac{\omega_e^2(1+k^2\lambda_D^2)}{\rho_0k^2} \int_{-\infty}^{\infty} dv\frac{\partial f_0(v)/\partial v}{v-\omega/k +i0}
\end{equation}
which here is extended to include an entropy-generated factor $(1+k^2\lambda_D^2)$.  The entropic effect in this case occurs at short wavelengths the order of the Debye length $\lambda_D$. Solution of $\mathcal{D}(k,\omega)=0$ is prescribed by the Landau contour in the complex $v$-plane to yield the collisionless Landau damping $\gamma_L(k)$. In  thermal plasma with Maxwellian distribution $f_0(v)$ it causes a minor modification of the Langmuir wave frequency
\begin{equation}
\omega_L\approx\pm\omega_{e} \big(1+2k^2\lambda_D^2\big)+i\gamma_L(k)%\\
%\gamma_L(k)&\approx&-\sqrt{\pi/8}\big(\omega_e/k^3\lambda_D^3\big)\exp\big[-{\textstyle{\frac{1}{2}}}(k\lambda_D)^{-2}-2\big]
\end{equation}
replacing the familiar ratio $\frac{3}{2}$  in the brackets by the integer 2, while leaving Landau damping about unaffected. Accounting for the entropy effect thus causes a slightly stronger increase of the frequency with wavenumber which also will slightly affect the wavenumber dependence of the thermal fluctuation spectrum. Similar minor changes can be expected in the dispersion relations of other plasma waves.

The collision term on the right in (\ref{eq-liou-3}), which in this Liouville-Vlasov approach is neglected, is in fact a rather complicated integral which in its simplest  binary collisional form gave rise to the definition of the Gibbs-Boltzmann entropy we used in the above. In a stricter theory, $\mathcal{C}$ is a huge extended sum over a long series of integrals  \citep{bogoliubov1962} accounting for all higher interactions between increasing numbers $N$ of particles involved. It is so far not known what effects the inclusion of the $N$-particle entropy $S_N=\log f_N(\vec{x}_N,\vec{v}_N,\vec{x}_{N-1},\vec{v}_{N-1}, \dots\vec{x}_1,\vec{v}_1) $ would cause. 

\section{Entropy-kinetic equation}
Remaining on the level of the above used one-particle phase-space distribution $f_1(\vec{x},\vec{v},t)\equiv f$, and referring to Gibbs-Boltzmann's kinetic definition of the local entropy $S=S_B=\log f$ at a given phase-space point $(\vec{x},\vec{v})$, an evolution equation for the kinetic entropy $S(\vec{x},\vec{v},t)$ can be derived directly when dividing the one-particle Liouville equation by the one-particle distribution function $f(\vec{x},\vec{v},t)$. Since the Hamiltonian $\mathcal{H}$ is a linear operator, this generates a kinetic equation for $S$, the entropy itself:
\begin{equation}
\partial_t\,S+[\mathcal{H}_0,S]_P+(\nabla\,\phi_s[S])\cdot\partial_{\vec{p}}S =\mathcal{C}/f
\end{equation}
(written in terms of the momentum $\vec{p}$ instead velocity $\vec{v}$) which can be understood as the wanted one-particle evolution equation of the kinetic Gibbs-Boltzmann entropy $S$ in the underlying one-particle phase space $(\vec{p},\vec{x})$.\footnote{It should be noted here that we prefer using Boltzmann's kinetic definition of entropy. Gibbs definition as the sum of the probability-weighted Boltzmann entropies $\int d\vec{p}\ f\log f$ refers  to the canonical level. Though it is also possible to derive a kinetic equation for the Gibbs entropy-density $f\log f$ in phase space  when multiplying the Liouville equation by $1+\log f$, little is gained then instead some more severe complications with the collision term.} Here $\phi_s[S]$ is the above given entropic force-functional as integral over the chosen relevant momentum-space $V_{\vec{p}}=\int_{\Delta V_{\vec{p}}} d\vec{p}$. It thus contains the total entropy the momentum-space volume contributes and, in this way, reacts on itself through the last term. The right-hand side of this equation is well defined for any normalized non-zero distribution function $f\neq0$, the normal case of course. As usual, the Hamilton equations of motion are implicit to this expression and of course contain/require the prescription of the dynamics. Spelled out in momentum $\vec{p}$ instead velocity, the last equation becomes explicitly
\begin{equation}\label{eq-kin-ent}
\partial_t\,S(t,\vec{x},\vec{p})+\frac{\vec{p}}{m}\cdot\nabla\,S(t,\vec{x},\vec{p})+\dot{\vec{p}}\cdot\partial_{\vec{p}}S(t,\vec{x},\vec{p})=\mathcal{C}\mathrm{e}^{-S(t,\vec{x},\vec{p})}\equiv\mathcal{C}'
\end{equation}
where the force term $\dot{\vec{p}}=\vec{F}([S],\vec{x},\vec{p},t)$ in the equation contains the external forces and includes the internal entropic force, and on the right we have taken advantage of the definition of the kinetic Gibbs-Boltzmann entropy to replace the one-particle kinetic distribution function $f=\exp S$ through it. (One may remember that this means that the local phase-space probability-density is proportional to the local phase-space volume $f\propto\Omega=\exp S$.) This replacement also applies to the collision term in order to close the equation as a function of $S$ alone. The same replacement must also be done in any of the dynamical field equations which govern the dynamics of the particles and hence also the dynamics of the entropy. 

In this sense the above kinetic equation completely describes the generation and phase-space evolution of entropy (or information) under the considered Hamiltonian dynamical process $\mathcal{H}(\vec{x},\vec{p},[S])$ with the accessible volume-entropy (or information) to be obtained as its Gibbs-Boltzmann moment $\Delta S(\vec{x},t)=\int_{V_{\vec{p}}} d\vec{p}\, f\log f$. It is clear from here that taking the moment of this kinetic equation one arrives at an average or even fluid equation for the evolution of entropy in real space-time under general non-equilibrium conditions.

In a more general sense, one can consider Eq. (\ref{eq-kin-ent}) as a fundamental equation in physics and probably also, in greater generality, as a fundamental equation governing the dynamics of information and information transport affected by the extraordinarily complicated collision term on the right and by entropy/information itself. 

There are two  difficulties to overcome when attempting to solve the kinetic entropy equation in application to real problems. The first is that, though being fundamental to the evolution of entropy, Eq. (\ref{eq-kin-ent}) is an \emph{integro-differential} equation through the entropic force acting in phase space. The second difficulty lies in the complicated collision term, which is defined through higher-order interactions and thus through higher-order entropies. To lowest order in the local deviation of entropy the inconvenient exponential can be cautiously expanded and replaced by\footnote{One would write $S=S_0+\delta\,S$. However, $S_0=$ const, the total entropy in the global volume plays no role. So we can either  retain it which in the exponential expansion just produces the exponential e$^{S_0}$ and normalizes $\delta S/S_0$. Or, since $S_0$ is an unmeasurable constant, we can also put $S_0=1$ with the entropy change $\delta S$ in all expressions understood as being re-normalized.}  
\begin{equation}
\exp\,S\approx \mathrm{e}^{S_0}\big(1+\delta\,S/S_0\big)
\end{equation}
 to be used conveniently in the underlying Hamiltonian and the field equations.

The philosophy behind this approach is the usual probabilistic picture of defining the probability density in a certain small though finite volume $\Delta V_{\vec{p},\vec{x}}=(d\vec{p}\,d\vec{x})$ of phase space. This volume may be small but must contain sufficiently many particles/constituents at any time to be integrated over, in order to be able to define a probability density $f(\vec{p},\vec{x},t)$ at a given time $t$ (a sufficiently short time interval $\Delta t$ prescribed by the time resolution of measurement). Considering the equivalent picture drawn in chaos theory which deals with the nonlinear interaction between just a few particles and their dependence on the initial conditions and measurement errors, so in the statistical Liouvillian phase-space approach, these interactions are averaged over; they are sub-kinetic (microscopic) to it and contribute only in as far as they leave a trace or signature on the probability in the kinetic phase space density $\Omega\propto f$. Since the Hamiltonian in principle contains all these interactions, it takes them  into account  in the size of the phase space volume, at least in an average sense. Their sub-microscopic (chaotic) resolution would require to solve the $N$-particle phase space Liouville equation for all the $N-1$ particles interacting with the $N$th particle, which includes all the $N-1$ single-particle phase-space volumina including all their mutual interactions. This problem can be solved only by coarse graining and thus only approximately. It shows however that chaos theory without application of such a coarse graining procedure describes reality only if it considers a countably and manageably small number of particles such that a reasonable particle \emph{density} for the phase-space unit can be defined, not the rare case of a collection of a small number of single mutually interacting particles. Poincar\'e-chaos theory describes reality on the sub-kinetic level where it hardly is accessible to measurement in any system that is built up of very many constituents, that is of most real systems. The chaotic approach of few particles only is from this point of view to be considered as most interesting but below accessible reality in most physical problems if not leading to observable kinetic effects whose detection requires reference to renormalization-group techniques \citep{wilson1971}.  

The overall  integrated entropy $\langle S\rangle=\int d\vec{x}d\vec{p}\, f\log f$ under stationary conditions is of little interest as  it is just a number, a huge constant of integration and in any dynamical process can be neglected. The entropy-kinetic equation can be reduced to an equation for the entropy fluctuations in momentum space $\delta S$, allowing for functional linearization 
\begin{equation}
\partial_t\delta S+\big(\vec{p}/m\big)\cdot\nabla\delta S+\big\{\vec{F}_0+\delta\vec{F}[S]\big\}\cdot\partial_{\vec{p}}[S+\delta S]=\delta\mathcal{C}'
\end{equation}
which in this form accounts for the local variation of the entropy while still containing all the higher order products of the variation $\delta S$. Here $\vec{F}_0$ is the external force involving the fields, and $\delta\vec{F}$ is the variation of the force with respect to the entropy. The last term  maintains the stationary spatially homogeneous entropy $S(\vec{p})$ because, as a function of momentum $\vec{p}$ like, for instance under stationary and homogeneous conditions it gives $\partial_t S_B=\nabla S_B= 0$, but $\partial_{\vec{p}}  S_B(\vec{p})\neq0$. 

Formally the kinetic equation is an  identity (the total time derivative) of similar kind as the Liouville equation. It describes the phase-space evolution of the one-particle entropy $S(\vec{x},\vec{p},t)\equiv S_1(\vec{x}_1,\vec{p}_1,t)$ under the dynamical action imposed by the Hamiltonian $\mathcal{H}_0$ which contains the involved fields (for instance the electromagnetic field). In this general form, it is  so far restricted just to the one-particle entropy $S$ and the one-particle distribution function $f_1=f$ but could formally become generalized to the $N$-particle case (developed in \citep{bogoliubov1962} and somewhere else, the famous BBGKY hierarchy). The space-dependent expectation value of the entropy (which enters the entropic force) is obtained, in the usual way, as the statistical momentum-space average
\begin{eqnarray}
\langle S(\vec{x},t)\rangle &=&\int_{\Delta V_{\vec{p}}} d\vec{p}\,f(\vec{x},\vec{p},t)\, S(\vec{x},\vec{p},t) \cr
&=&\int_{\Delta V_{\vec{p}}} d\vec{p}\, S(\vec{x},\vec{p},t)\exp[S(\vec{x},\vec{p},t)]\,
\end{eqnarray}
The integration is over the sufficiently large momentum-space volume $\Delta V_{\vec{p}}$. Given an underlying dynamics, it yields the momentum-space average of the entropy $S(\vec{x},\vec{p},t)$ with retained dependence on configuration space. Once again this expression demonstrates the self-determination of the entropy. 

The entropy in this representation just exchanges its place in the integral, becoming its own probability distribution which weighs its inverse dependence, the exponential of itself. In this form the entropy appears as the momentum-space probability which weighs the kinetic phase space volume element $\Omega=\mathrm{e}^S$. Interpreted that way, it says that the \emph{entropy controls the evolution of the phase space volume} occupied by Hamiltonian dynamics. The entropy affects its hosting phase space volume $\Omega$. Locally however this affection does not simply proceed in a smooth expansion of phase space as this is the product of momentum and configuration spaces.  

Given the importance of information in the modern world, an equation like the one derived here from fundamental considerations including the entropic force should have some  application and thus could be of practical use.  

Having derived the kinetic evolution equation for the kinetic entropy, a most interesting problem arises when asking for the second law and its maintenance on the kinetic level. The second law offers a constraint on the entropy evolution which has to be incorporated into a full theory. It requires discussing what, on the kinetic level, the notion of an \emph{exclusively growing} entropy means and how it has to be understood or re-interpreted. Clearly, the \emph{global over-all} entropy in configuration space $S_0(t)=\int d\vec{v}d\vec{x}\, f\log f$ will only grow at $\Delta S_0(t)>0$. The same should also hold for any \emph{isolated} spatial volume $\Delta V$. However, in momentum space permanent increase of $S$ is clearly not an obligation as energy and heat can be distributed with respect to the moments $\vec{p}$ in phase space. Here we do not enter into a deeper investigation of the related questions but relegate it instead to an own consideration.   

This derivation of the kinetic entropy equation remains in the realm of classical physics.\footnote{A quantum approach is beyond this work but should make use of von Neumann entropy.} There are two basic classical theories, electrodynamics and general relativity where it could be applied. In electrodynamics this does not look impossible as the Hamiltonian of the underlying charged particles is known. We have demonstrated its effect on the example of Langmuir waves above. General relativity, on the other hand, is not a particle but a field theory to which our particle approach does primarily not apply. Its kinetic consideration should refer to some non-quantum version of  \emph{Geometrodynamics} but requires the field-theoretical version of kinetic theory which is beyond the present investigation. 

The electrodynamic field is a gauge field $\vec{A}$ which enters the generalized momentum $\vec{p}'=\vec{p}\pm ie\vec{A}$, where $e$ is the elementary charge. It is to be accomplished with the field equations including sources, charge and current. Replacing the distribution function $f_\pm=\exp S_\pm$ it becomes formally possible to rewrite the kinetic theory in terms of the entropies $S_\pm$ under the action of the electrodynamic gauge field $\vec{A}$. Since the field equations couple the entropy distributions of the positive and negative charges, it is clear that, similar to ordinary kinetic theory, separate though coupled kinetic equations apply to $S_\pm$. Ultimately, in electrodynamics (plasma physics) the relevant entropy at location $\vec{x}$ in real space  is obtained as the sum
\begin{equation}
 S_\mathit{tot}(\vec{x},t)=\sum_{+,-} \int_{\Delta V_{{p}}} d\vec{p}\ S_\pm(\vec{x},\vec{p},t) \exp {S_\pm}(\vec{x},\vec{p},t)
\end{equation}

It is obvious that, in principle, it is the entropy which on the kinetic level determines the dynamics of the system, with the probability distribution $f(\vec{x},\vec{p},t)$ having occurred in an intermediate and incomplete step. It will be interesting to infer which consequences this insight might have in quantum physics, where one probably would need to refer to von Neumann instead Gibbs-Boltzmann entropy, and the density matrix instead the classical distribution function. 

%As a final remark we refer to the repulsive nature of the entropic force as used in this investigation. If it will become possible to formulate an entropic kinetic field theory along the lines prescribed here, application to the classical gravitational field would be of vital interest. The gravitational entropic force in that case should provide a gravitational self-repulsion which cosmologically could be understood causing the expansion of the universe. Similarly, in quantum field theory application to the Higgs field could lead to interesting consequences.  

\section{Summary}
The present note started from the definition of a classical non-quantum entropic force as was proposed in \citep{treumann2019a}. This force acts like any mechanical force on particles or fluids. We gave it an explicit expression for a fluid system and applied it in passing to magnetohydrodynamics where it, in linear theory, causes just small modifications in the magnetosonic wave dispersion relation. Applied to magnetohydrodynamic turbulence it leads to a modified inertial range spectral slope on the transverse velocity fluctuation spectrum which might be remarkable just in the magnetohydrodynamic ion-inertial range.  A more elaborate theory might be useful to be put forward in the context of electron magnetohydrodynamics \citep{gordeev1994,lyutikov2013} when applying the same or similar reasoning there. Numerical electron-MHD simulations favourably reproduce the proposed spectral shape.

We then proceeded to the more fundamental question of the entropic force in kinetic theory. This becomes a rather involved problem. However, skipping those difficulties, we chose to attempt the derivation of a kinetic equation for Gibbs-Boltzmann's entropy, not the phase space distribution function, following Gibbs-Boltzmann's philosophy that the entropy is the logarithm of the distribution function  in classical physics. The equation obtained is the phase-space evolution equation of the kinetic entropy under the action of a given Hamiltonian which includes the entropic force-potential. 

This equation should provide the basis of the evolution, self-interaction and,  possibly, also the self-generation of entropy (or information) in any classical Hamiltonian system. Such a self-generation would arise if taking into account the self-consistent version of the collision term $\mathcal{C}$ on the right of the equation as this includes all  the nonlinear interactions in an $N$-dimensional phase space as suggested in BBGKY theory \citep{bogoliubov1962} of classical many-particle kinetic theory. A similar quantum approach in many-body theory \citep{fetter1971} will probably face severe hurdles. Nevertheless, it awaits its construction.   
 %\vspace{1cm}

%\section*{Author Contributions}
%WB: discussion, redaction. RT: conception, analysis, writing.

%\section*{Funding}
%This work was not funded by any external sources.

%\section*{Acknowledgments}
%This work was part of a brief Visiting Scientist Programme  (by RAT) at the International Space Science Institute Bern. We acknowledge the interest of the ISSI directorate as well as the hospitality of the ISSI staff, in particular the assistance of the librarians Andrea Fischer and Irmela Schweitzer, and the Systems Administrator Willi W\"afler. 

\end{document}